\documentclass[superscriptaddress,pra,showpacs,floatfix]{revtex4}

\usepackage{amsmath}
\usepackage{amssymb}
\usepackage{amsfonts}
\usepackage{pifont}
\usepackage{colordvi}
\usepackage{color}
\usepackage{bm}
\usepackage{bbm}
\usepackage{epsf}
\usepackage{epsfig}
\usepackage{nicefrac}
\usepackage{graphicx}
\usepackage{dcolumn}
\def\half{{\textstyle{1\over 2}}}

\def\ii{{\mathrm{i}}}
\def\dd{{\mathrm{d}}}

\begin{document}

\newcommand{\borel}[1]{\mbox{$\mathcal{#1}$}}

\newcommand{\addrHD}{Max--Planck--Institut f\"ur Kernphysik,
Postfach 103980, 69029 Heidelberg, Germany}

\newcommand{\addrHDphil}{Institut f\"ur Theoretische Physik,
Universit\"{a}t Heidelberg,
Philosophenweg 16, 69120 Heidelberg, Germany}

\newcommand{\addrDAPNIA}{DAPNIA, Commissariat 
\`{a} l'\'{E}nergie Atomique,
Centre de Saclay, 91191 Gif--Sur--Yvette, France}

\title{Structure, Time Propagation and Dissipative Terms for Resonances}

\author{U. D.~Jentschura}
\affiliation{\addrHD}
\affiliation{\addrHDphil}
\author{A. Surzhykov}
\affiliation{\addrHD}
\author{M. Lubasch}
\affiliation{\addrHDphil}
\author{J. Zinn--Justin}
\affiliation{\addrDAPNIA}

\date{\today}

\begin{abstract}
For odd anharmonic oscillators, it is well known that
complex scaling can be used to determine 
resonance energy eigenvalues and the corresponding 
eigenvectors in complex rotated space. 
We briefly review and discuss various methods
for the numerical determination of such eigenvalues, 
and also discuss the connection to the 
case of purely imaginary coupling, which is 
${\mathcal{PT}}$-symmetric.
Moreover, we show that a suitable generalization of 
the complex scaling method leads to an 
algorithm for the time propagation of wave packets in potentials
which give rise to unstable resonances.
This leads to a certain unification 
of the structure and the dynamics.
Our time propagation results agree with 
known quantum dynamics solvers
and allow for a natural incorporation of structural 
perturbations (e.g., due to dissipative processes) into the 
quantum dynamics.
\end{abstract}

% 11.15.Bt General properties of perturbation theory
% 11.10.Jj Asymptotic problems and properties
% 68.65.Hb Quantum dots
% 03.67.Lx Quantum computation
% 85.25.Cp Josephson devices

\maketitle

\section{INTRODUCTION}
\label{intro}

Quantum tunneling is one of the intriguing phenomena 
in quantum mechanics. Since the seminal work of G.~Gamow 
on the theory of the alpha decay of a nucleus~\cite{Ga1929}, 
quantum tunneling of a particle trapped in a metastable 
potential well has been studied in detail in many areas of 
physics. Indeed, in mathematical terms, the decay width 
of the corresponding resonance state in an, e.g., 
cubic potential can be traced back 
to an instanton configuration which is a solution of 
the classical equations of motion of a particle 
moving in the ``inverted'' potential $-V(q)$ in such a way that the 
classical action of the particle along its trajectory 
remains finite even though the time domain covered 
by its trajectory is the entire space $\mathbbm{R}$.
Here, the instanton configuration covers the 
domain naturally associated with the tunneling 
process from a relative minimum 
of the potential to a point where a horizontal line 
would emerge from the ``tunnel'' potential.
An approximation to the width can be obtained by considering
the fluctuations around the classical path, and by 
a subsequent evaluation
of the Fredholm determinant describing the fluctuations. 
This is quite analogous to the case of double-well-like 
potentials~\cite{ZJJe2004i,ZJJe2004ii}, where the instanton 
configuration describes an oscillatory motion 
covering a trajectory oscillating between the 
two degenerate maxima of the inverted potential.

At the same time, the structure of the potentials depends 
very much on the complex phase of the coupling constant. 
Let us assume a metastable potential 
approximated by a one-dimensional
harmonic oscillator perturbed by a cubic term 
$\sqrt{g} \, q^3$, where $g$ is the coupling constant. 
For purely imaginary $\sqrt{g} = \ii \beta$, with 
$\beta \in \mathbbm{R}$, the Hamiltonian is invariant 
under the composed application of a parity transformation 
and a time reversal and thus called $\mathcal{PT}$-symmetric.
In some sense, the $\mathcal{PT}$-symmetric case is 
a natural generalization of hermiticity (or even self-adjointness) 
to the complex domain, and it will be verified here that a number
of concepts known to be applicable to 
``stable'' anharmonic oscillators are applicable to the 
$\mathcal{PT}$-symmetric cubic potential as well.

Coming back to the case of real coupling parameter $\sqrt{g}$, 
we note that known 
methods for the numerical determination of resonance energies
and widths include the diagonalization of a complex rotated 
Hamiltonian matrix, Borel 
resummation~\cite{BorelPhi4,BeWu1969,BeWu1973} in complex directions 
of the parameters, and strong-coupling expansions.
These three methods will be discussed and contrasted here as
they are important for the construction of an adiabatic, 
complex transformed time propagation algorithm that is also
described in this article. Namely, we attempt to solve 
the problem of how to propagate a wave packet that 
moves under the influence of a potential with metastable 
resonances and which may thus even ``escape'' to the 
classical region of attraction where the potential 
assumes large negative values, without the need for a 
temporally adjusted numerical grid and 
without the need for the introduction
of transparent boundary conditions. Moreover, we attempt to 
construct this algorithm using complex resonance
eigenstates, thus giving a manifest interpretation to
the complex resonance state and energies within the time propagation 
method, including the back-transformation of the complex
rotated and propagated states to the normal coordinate representation.

This paper is organized as follows. In Sec.~\ref{theory},
we recall some basic definitions related to the cubic
potential, to perturbative and strong-coupling expansions
and to the method of complex scaling. 
Also, corresponding numerical investigations are discussed.
In Sec.~\ref{dynamics}, 
we discuss a method for time propagation
in the cubic potential, which leads to the 
above mentioned desired unification.
Finally, a summary is presented in Sec.~\ref{summary}.
An appendix is devoted to the discussion of potential 
applications of the algorithms discussed here 
within a solid-state physics context.
In the entire article, we attempt to 
follow a rather detailed style in the presentation and 
hope that the reader will not find the level of detail
excessive.

%
% RESONANCE ENERGIES
%
\section{COMPLEX RESONANCE ENERGIES VERSUS REAL
$\boldsymbol{\mathcal{PT}}$--SYMMETRIC SPECTRUM FOR THE CUBIC OSCILLATOR}
\label{theory}

%
% Orientation
%
\subsection{Orientation}
\label{theoryA}

Although our analysis can easily be generalized to 
quintic and other ``odd'' potentials with metastable resonances, 
we start for simplicity with the one--dimensional Hamiltonian 
of the cubic potential in the normalization
\begin{equation}
\label{GENcubic}
H_{\rm c} \, = \, -\frac{1}{2} \,
\frac{\partial^2}{\partial q^2}  + \frac{1}{2} \, q^2 + 
\sqrt{g} \, q^3 \,,
\end{equation}
which recovers the harmonic oscillator spectrum
$E_N = N + \half$ in the limit of a vanishing
coupling constant $g \to 0$, where $N$ is a nonnegative integer and 
constitutes the quantum number of the state.
For nonvanishing $g$, the operator~(\ref{GENcubic}) is
not self--adjoint, and it does not possess a spectrum of discrete, real
energy eigenvalues.
For real and positive~$g$, the 
cubic Hamiltonian (\ref{GENcubic}) 
possesses resonances, i.e.~poles of the resolvent
$(H_{\rm c} - E)^{-1}$ which is {\em a priori}
defined as a holomorphic function 
for ${\rm Im}\, E > 0$. The poles (resonances) become apparent in the
meromorphic analytic continuation of the resolvent
to ${\rm Im}\, E < 0$. The resonance energy eigenvalues
are of the form
\begin{equation}
\label{energy_cubic_oscillator}
E_{N}(g) = {\rm Re}[E_{N}(g)] - \ii \, \frac{\Gamma_{N}(g)}{2} \,.
\end{equation}
By contrast, for purely 
imaginary coupling $\sqrt{g} = {\rm i}\,\beta$, the Hamiltonian
(\ref{GENcubic}) is ${\mathcal PT}$-symmetric,
i.e.~it is invariant under a simultaneous parity
transformation $q \to - q$ and a time reversal
operation $t \to -t$, the latter being equivalent to
an explicit complex conjugation of the Hamiltonian and thus
to the replacement $\beta \to -\beta$.

Based on numerical evidence, it has been conjectured
around 1985 by D.~Bessis and one of us (J. Z.--J.) in
private communications that
the spectrum of the ${\mathcal PT}$-symmetric Hamiltonians
of odd anharmonic oscillators should consist of real eigenvalues,
even if these Hamiltonians are obviously not self-adjoint.
C.~M.~Bender and others have recently
studied ${\mathcal PT}$-symmetric Hamiltonians quite
intensively 
(see, e.g., Refs.~\cite{BeBo1998,BeBoMe1999,BeWe2001,BeDuMeSi2001,BeBrJo2002}).
Note, in particular,
that even the quartic anharmonic oscillator with negative
coupling becomes a ${\mathcal PT}$-symmetric Hamiltonian
with real spectrum when endowed with appropriate boundary
conditions imposed on the wave function as a function of a
complex coordinate, as detailed in Eq.~(4) of
Ref.~\cite{BeBoJoMeSi2001}.  
The conjecture has recently been supported on
mathematical grounds
(see~Refs.~\cite{Sh2002cmp,DoDuTa2001i,DoDuTa2001ii,DoDuTa2005}).
Important further contributions to the development of the 
theory of ${\mathcal PT}$-symmetric quantum mechanics have
been summarized in Refs.~\cite{BeBo1998,BeBoMe1999,BeBrJo2002}.

%
% Figure 1 
%
\begin{figure}[t]
\begin{center}
\begin{minipage}{12cm}
\begin{center}
\vspace*{-0.0cm}
\includegraphics[width=0.6\linewidth,angle=0, clip=]{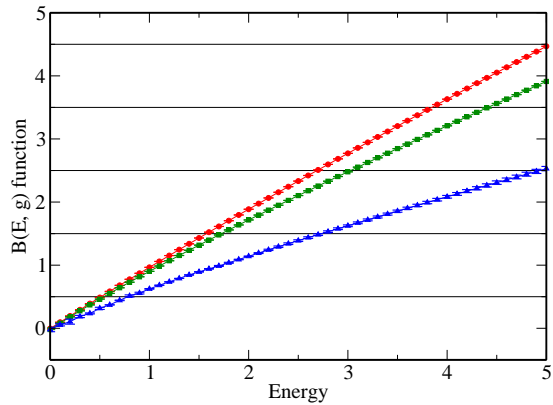}
\vspace*{-0.0cm}
\caption{\label{Fig1} (color online.) 
Energy dependence of the function $B(E, g)$
for the $\mathcal{PT}$-symmetric case.
Results are presented for the coupling 
parameters $\sqrt{g}= {\rm i}\beta$
with $\beta = 1/10$ (red circles), 
$\beta = 1/5$ (green squares) and $\beta = 1$ (blue triangles).}
\end{center}
\end{minipage}
\end{center}
\end{figure}
%
%

%
% Table 1
%
\begin{table}[t]
\begin{center}
\begin{minipage}{12cm}
\begin{center}
\caption{\label{table3} Ground-state energy of the cubic 
Hamiltonian for the $\mathcal{PT}$-symmetric case
$\sqrt{g} = {\rm i}\,\beta$. 
Results have been obtained by the diagonalization of the
complex-transformed Hamiltonian in the basis of the harmonic oscillator 
wavefunctions (``exact'' values) and by solving the 
quantization condition given in Eq.~(\ref{BSQC}).
The apparently ``converged'' decimals for the solutions of $B(E, g) = \half$
are underlined.}
\vspace*{0.3cm}
\begin{tabular}{l@{\hspace{0.5cm}}l@{\hspace{0.5cm}}l} 
\hline
\hline
\rule[-3mm]{0mm}{8mm}
$\beta \, \, \, $   
& $E_{0}(g)$ & Solution of $B(E, g) = \half$ \\
\hline
\rule[-3mm]{0mm}{8mm}
$1/10$ & 0.512~538~145 & \underline{0.512~538~145}     \\
\rule[-3mm]{0mm}{8mm}
$1/8$ &  0.518~760~345 & \underline{0.518~760~34}4(1)  \\
\rule[-3mm]{0mm}{8mm}
$1/6$ & 0.530~781~759  & \underline{0.530~781~7}7(1) \\
\rule[-3mm]{0mm}{8mm}
$1/4$ & 0.558~372~124  & \underline{0.558~37}7(4)\\
\rule[-3mm]{0mm}{8mm}
$1/2$ & 0.645~877~080  & \underline{0.645}~7(3) \\
\hline
\hline
\end{tabular}
\end{center}
\end{minipage}
\end{center}
\end{table}
%
%

%
% Case of $\boldsymbol{\mathcal PT}$ symmetry
%
\subsection{Imaginary coupling parameter and real energies}
\label{theoryB}

The discussion in this section is {\em a priori} relevant 
only for the case $\sqrt{g} = \ii \beta$, with 
$\beta \in \mathbbm{R}$, but we will keep a general $g$
in the first considerations, because the general 
formulas (as a function of $g$) will
be useful later in this article.
First, we recall that, 
as explained in Refs.~\cite{ZJJe2004i,ZJJe2004ii},
the spectrum of anharmonic oscillators can be described
in many cases by two functions $B$ and $A$.
Respectively, these are
related to the perturbative expansion about the
minima of the potential and to the tunneling of the
quantal particle from one degenerate or
quasi-degenerate minimum (if it exists) to the
other degenerate or
quasi-degenerate minimum of the potential.
Specifically, around Eq.~(3.31) of
Sec.~3 of Ref.~\cite{ZJJe2004i},
it is explained how, in the case of integrable 
systems, a ``perturbative $B$ function'' can be
obtained from the perturbative
expansion of the logarithmic derivative of the
wave function, which fulfills a differential
equation of the Riccati type and which is subject
to a uniqueness condition which gives rise to the
integer $N$ on the right-hand side of Eq.~(\ref{BSQCpre}).
Using techniques outlined in Ref.~\cite{ZJ1984jmp},
the function $B(E, g)$ can be easily determined 
for the cubic potential, and the first terms read as follows,
\begin{eqnarray}
\label{B_series}
B(E, g) &=& E + g \left( \frac{7}{16} + 
\frac{15}{4} \, E^2 \right) 
+ g^2 \left( \frac{1365}{64} \, E + \frac{1155}{16} \, E^3 \right) +
\mathcal{O}(g^3) \, .
\end{eqnarray}
By formulating the initial perturbation as $\sqrt{g}\, q^3$, we have
obtained a perturbative expansion for $B$ which involves integer
powers in the coupling constant. As usual, the perturbative
expansion for the $N$th energy level can be obtained 
by inverting the condition
\begin{equation}
\label{BSQCpre}
B(E, g) = N + \half \, ,
\end{equation}
and we obtain the standard Rayleigh--Sch\"{o}dinger
perturbation theory (RSPT) series for the $N$th 
level of the cubic potential,
\begin{equation}
\label{RScubic}
E_{N}(g) = \sum\limits_{k = 0}^{\infty} E^{(0)}_{N,K} \, g^k \, ,
\end{equation}
where $K$ is the order of perturbation theory, and
the leading perturbative coefficients read as follows,
for a general level with quantum number $N$,
\begin{subequations}
\label{pertcoeff}
\begin{align}
E^{(0)}_{N,0} =& \; N + \half \,,
\\[1ex]
E^{(0)}_{N,1} =& \; - \left[ \frac{7}{16} + 
\frac{15}{4} \left( N + \half \right)^2 \right] \,, 
\\[1ex]
E^{(0)}_{N,2} =& \; - \left[ \frac{1155}{64} \left( N + \half \right) +
\frac{705}{16} \left( N + \half \right)^3  \right] \,, 
\\[1ex]
E^{(0)}_{N,3} =& \; - \left[ \frac{101479}{2048} 
+ \frac{209055}{256} \left( N + \half \right)^2 
+ \frac{115755}{128} \left( N + \half \right)^4 
\right] \,.
\end{align}
\end{subequations}
The minus signs are explicitly indicated to illustrate
that all perturbative coefficients are negative 
for all levels, except for the leading term $N + \half$, 
which stems from the unperturbed harmonic oscillator.

If perturbation theory determines the energy eigenvalues 
completely in the $\mathcal{PT}$-symmetric case
and there are no instanton configurations to 
consider, then the quantization condition (\ref{BSQCpre})
can be reformulated as
\begin{equation}
\label{BSQC}
\frac{1}{\Gamma\left(\half - B(E, g) \right)} = 0 \,, \qquad
\sqrt{g} = \ii\, \beta.
\end{equation}
This quantization condition is formulated such as to 
display the analogy to those for more complex potentials like 
the double well [see Eq.~(2) of Ref.~\cite{JeZJ2001}], but it lacks 
the ``instanton $A$ function'' present in the cited equation.
A Bohr--Sommerfeld quantization condition of the form 
(\ref{BSQC}) is relevant 
for stable anharmonic oscillators like, e.g., the 
quartic one with a perturbation proportional 
to $g \, q^4$ for positive coupling $g$, where there are 
no instanton configurations to consider and 
therefore no ``instanton $A$ function'' present.
We now investigate to which extent this quantization condition
could be relevant for the cubic anharmonic oscillator.

To this end, we first recall that 
the cubic Hamiltonian (\ref{GENcubic}) 
displays ${\mathcal PT}$-symmetry for imaginary coupling 
$\sqrt{g}= {\rm i}\beta$. The spectrum of the 
cubic oscillator becomes real in this case, as a 
consequence of pseudo--Hermiticity~\cite{Mo2002}, and the 
perturbation series (\ref{RScubic}) becomes an alternating, factorially 
divergent series in the variable 
$g = - \beta^2$; series of this type are typically 
Borel-summable. We are therefore led to the conjecture that the quantization 
condition (\ref{BSQC}) should describe the energy levels of the 
cubic Hamiltonian (\ref{GENcubic}), 
but only for the ${\mathcal PT}$-symmetric 
case $\sqrt{g}= {\rm i}\beta$.

As discussed in Sec.~\ref{theory} and 
as shown in our previous paper (Ref.~\cite{SuLuZJJe2006}), 
the direct analysis of the perturbative $B$ function
which enters the quantization condition~(\ref{BSQC}), 
without any detour via the perturbation
series~(\ref{RScubic}), can be used to
investigate our conjecture formulated above, in order to 
calculate the energy eigenvalues of the 
cubic Hamiltonian (\ref{GENcubic}) directly.
Here, we explore this approach in the 
case of imaginary coupling $\sqrt{g}= {\rm i}\beta$. 
To this end, we interpret $B(E, g)$ as a 
function of $E$ (at fixed $g$) and numerically 
determine the points at which 
this function assumes values of the form $N + \half$. According to 
Eq.~(\ref{BSQC}), these points correspond to the real energy 
eigenvalues $E_N(g)$ of the cubic Hamiltonian in the 
${\mathcal PT}$-symmetric case. 
In many cases, direct investigation of the perturbative $B$ function 
has proved to be numerically more stable than that of the 
Borel-resummed series (\ref{RScubic}) 
(see alse Ref.~\cite{SuLuZJJe2006}).

In Fig.~\ref{Fig1}, for example, we display $B(E, g)$ as a function 
of the (real) energy argument $E$ for different fixed values of the
coupling parameters $g$. Specifically, we consider
the cases $\sqrt{g}= {\rm i}\beta = {\rm i}/10$ (circles), 
$\sqrt{g}= {\rm i}\beta = {\rm i}/5$ (squares) and 
$\sqrt{g}= {\rm i}\beta = {\rm i}$ (triangles). The points where 
$B(E, g) = N + \half$ are clearly displayed.
The (real) ground-state energy $E_N(g)$ 
for the $\mathcal{P T}$-symmetric case
is presented in Table~\ref{table3} and compared to reference
data obtained via a diagonalization of the Hamiltonian matrix.
The ground-state energy is well reproduced at $\beta = 1/10$
(up to 9 decimal digits). For stronger coupling, there is a 
larger uncertainty in the determination of the ground--state
energy value $E_{N=0}(g)$ because
the power series (\ref{B_series}) is divergent for all nonvanishing $g$,
and, hence, resummation techniques are required for its
calculation. In all cases, the function $B(E, g)$ is 
calculated by means of a generalized Borel--P\'{a}de method,
similar to Ref.~\cite{SuLuZJJe2006}.  We observe, in higher 
(Borel) transformation orders, 
an oscillatory behaviour of the Borel integral, evaluated along
complex directions, for larger $g$. These oscillations
cannot be overcome when the 
transformation order is increased and represent 
a fundamental limit of the convergence of resummed weak-coupling 
perturbation theory in the case of a large (modulus of the) coupling
parameter $g$. All numerical 
experiments support the conjecture (\ref{BSQC}) for the 
$\mathcal{PT}$-symmetric case.

%
% Figure 2 
%
\begin{figure}[t]
\begin{center}
\begin{minipage}{12cm}
\begin{center}
\vspace*{-0.0cm}
\includegraphics[width=0.9\linewidth,angle=0, clip=]%
{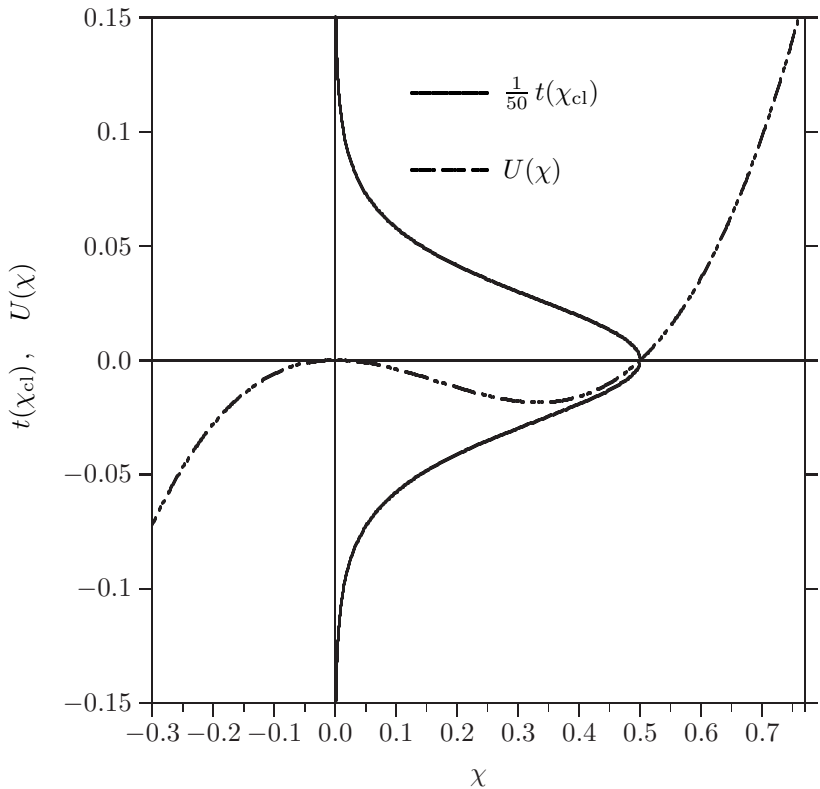}
\vspace*{-0.0cm}
\caption{\label{Fig2} Instanton configuration for the 
cubic potential. The dashed line is the 
potential $U(\chi) = \chi^3 - \half \, \chi^2$.
The solid line is the worldline of the 
instanton configuration 
$\chi_{\rm cl}(t) = [\cosh(t) + 1]^{-1}$, 
which reads in inverted form 
$t(\chi_{\rm cl}) = 
\pm {\rm arccosh}[(1-\chi_{\rm cl})/\chi_{\rm cl}]$.}
\end{center}
\end{minipage}
\end{center}
\end{figure}

%
% COMPLEX RESONANCE ENERGIES
%
\subsection{Real coupling parameter and complex resonance energies}
\label{theoryC}

We again consider the Hamiltonian (\ref{GENcubic}),
but this time for real and positive $g$. 
First, let us note that the structure of the 
perturbation series (\ref{RScubic}) is of course unaffected by a 
change in the complex phase of the coupling parameter.
However, we now have instanton configurations to consider.
Let us consider for a moment the classical Euclidean 
action, corresponding to (\ref{GENcubic}),
\begin{equation}
S[q(t)] = \int \dd t \, \left[ 
\frac12\, \left( \frac{\partial}{\partial t} q(t) \right)^2 + 
\frac12\, q(t)^2 + 
\sqrt{g}\, q(t)^3 \right] \,.
\end{equation}
This action describes the motion of a particle in the 
potential $-\sqrt{g}\, q(t)^3 - \half \, q(t)^2$.
Via the change of variable $q(t) = -\chi(t)/\sqrt{g}$, we obtain
the action,
\begin{equation}
S[\chi(t)] = \frac{1}{g} \, \int \dd t \, \left[ 
\frac12\, \left( \frac{\partial}{\partial t} \chi(t) \right)^2 + 
\frac12\, \chi(t)^2 - \chi(t)^3 \right] \,,
\end{equation}
for which the (redefined) potential now reads
$U(\chi) = \chi^3 - \half \, \chi^2$.
The (classical) instanton configuration is 
(see Fig.~\ref{Fig2})
\begin{equation}
\chi_{\rm cl}(t) = \frac{1}{\cosh(t) + 1} \,,
\qquad 
S[\chi_{\rm cl}(t)] = \frac{2}{15}\,,
\end{equation}
which fulfills $U(\chi_{\rm cl}(t=0)) = U(\half) = 0$.
An integration about the fluctuations around the instanton 
path in the partition function using methods described
in Ref.~\cite{ZJ1996} then leads to the known result
\begin{equation}
{\rm Im} \, E_0(g) \approx - \frac{1}{\sqrt{\pi \, g}} \,
\exp\left( - \frac{2}{15\,g} \right)
\end{equation}
for the imaginary part of the ground-state energy. 
Observe that in contrast to the quartic potential,
where two degenerate instanton configurations are
present because of the reflection symmetry of the 
quartic potential~\cite{ZJ1996}, the cubic potential has no reflection
symmetry and only one instanton.
We conclude that the ``instanton $A$ function'' for the 
cubic oscillator should have the leading terms
\begin{equation}
\label{instA}
A(E, g) = \frac{2}{15\,g} + \mathcal{O}(g) \,.
\end{equation}
However, we stress that the analogue of the quantization 
condition (\ref{BSQC}), suitably generalized for 
odd anharmonic potentials, has not yet appeared in the 
literature to the best of our knowledge.
This problem is currently under investigation.
It is clear that a suitable quantization condition should 
involve the $A$ function in such a way that the 
decay rate follows naturally by an expansion in both analytic
and nonanalytic terms for $g$ small, so that a 
so-called resurgent expansion~\cite{Ph1989,CaNoPh1993,Bo1994}
is obtained which allows for the nonanalytic behaviour 
of the imaginary part of the resonance energy as $g \to 0^+$. 

If the (generalized) quantization condition
for $g > 0$ were known, then we could use it in order
to calculate complex resonance energies, via a direct resummation
of both the perturbative $B$ and the instanton $A$ function,
in a similar way as was done for the $\mathcal{PT}$-symmetric 
case in the previous section. We recall that for the 
$\mathcal{PT}$-symmetric cubic potential, where we
resummed only the $B$ function, and for the 
Fokker--Planck potential (see Ref.~\cite{SuLuZJJe2006}), 
where we resummed both the 
$B$ as well as the $A$ function,
the direct resummation of the quantization condition
did not give a satisfactory numerical accuracy for the 
energy levels for moderate and large values of the 
coupling constant.
Here, our goal is to describe a numerical method which 
is applicable to all domains of the coupling constant,
and we thus continue here with a comparison of 
three methods for the calculation of resonance energies, 
including an evaluation of the 
domains of their respective applicability. 

{\em Method I.} We apply 
complex scaling (see Refs.~\cite{BaCo1971,YaEtAl1978}),
$q \to q \, {\rm e}^{{\rm i}\, \theta}$ 
to the cubic oscillator, which results in the 
Hamiltonian
\begin{eqnarray}
\label{GENcubic_scaled}
H_{\rm c}(\theta) &=& 
{\rm e}^{-2i\theta}
\left( - \frac{1}{2} \frac{\partial^2}{\partial q^2} + 
\frac12 \, q^2 \, {\rm e}^{4 \ii \theta} + 
\sqrt{g} \, q^3 {\rm e}^{5 \ii \theta} \right) \,.
\end{eqnarray}
The diagonalization of this complex scaled operator is
carried out in the basis of 
harmonic oscillator wavefunctions 
$\{ \phi_n(q) \}_{n = 0}^{N_{\rm max}}$,
for large enough $N_{\rm max}$, and the 
variation of the resonance energies under a suitable increase
of $N_{\rm max}$ is used to investigate the numerical
uncertainty of the results. This
allows us to numerically  determine the (complex) resonance energies of the 
original cubic Hamiltonian (\ref{GENcubic}). 
As discussed in Refs.~\cite{YaEtAl1978,Al1988}, 
these resonance energies are independent of $\theta$,
provided we choose $\theta$ sufficiently large
so that the rotated branch passes the position of the resonance 
under investigation.

{\em Method II.} 
We resum Rayleigh--Schr\"{o}dinger 
perturbation theory (RSPT) in complex directions of the parameters.
To this end,
we use the standard RSPT series for the cubic potential
as given in Eq.~(\ref{RScubic}).
Notice that the RSPT series is nonalternating in 
integer powers of $g$ for real $g$.
We then employ the Borel--Pad\'{e} summation method with the (Laplace) 
integration in the complex plane, as given by
Eqs.~(196)--(198) of Ref.~\cite{CaEtAl2007}.
The method has also been discussed in detail 
elsewhere~\cite{FrGrSi1985,Je2000prd,CaEtAl2007}, and it has
been put on rigorous mathematical grounds recently~\cite{Ca2000}
in the framework of distributional Borel summability. 
We employ an integration contour (called $C_{+1}$ in the conventions
of Ref.~\cite{CaEtAl2007}) which leads to a
negative sign of the imaginary part of the resonance energy 
eigenvalue, consistent with Eq.~(\ref{energy_cubic_oscillator}).
The accuracy obtainable using this weak-coupling method is restricted
by oscillations of the transforms
in higher orders, which are analogous to those reported in 
Ref.~\cite{SuLuZJJe2006} for other applications of the 
Borel--Pad\'{e} transforms.
Indeed, at a relatively moderate coupling $g = 0.6$, 
the ground-state energy as
determined by resummation cannot be calculated to better
accuracy than
\begin{equation}
E_0(g = 0.6) = 0.554(1)- 0.351(6) \,{\rm i}
\end{equation}
by resummation. Note that the oscillations cannot be overcome when the (Borel) 
transformation order is increased and represent 
a fundamental limit of the convergence of resummed weak-coupling 
perturbation theory in the case of a large (modulus of the) coupling
parameter $g$. A more accurate result obtained by method I for the 
same coupling parameter is
\begin{equation}
E_0(g = 0.6) = 0.554\,053\,519 - 0.351\,401\,778 \,{\rm i}\,.
\end{equation}

\begin{figure}[t]
\begin{center}
\begin{minipage}{10cm}
\begin{center}
\vspace*{-0.0cm}
\includegraphics[width=0.9\linewidth,angle=0, clip=]%
{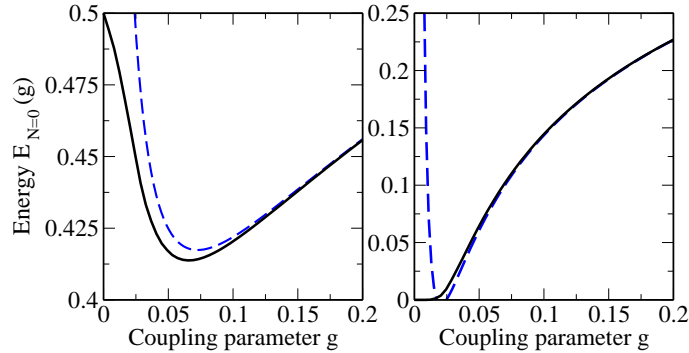}
\vspace*{-0.0cm}
\caption{\label{Fig3} (color online.) 
The real (left panel) and the imaginary part 
(right panel) of the exact resonance energy of the ground 
state for the cubic potential are displayed as a function of $g$
(solid lines). The real part approaches the values $\half$
as $g \to 0$, whereas the imaginary part vanishes in this limit.
The exact numerical values are compared to the sum of the 
first four leading terms of the strong-coupling asymptotics 
(dashed lines) for large coupling parameter $g$. The strong-coupling 
asymptotics are defined in Eq.~(\ref{expansion_higher_orders}),
and the coefficients are listed in Table~\ref{table1}. The first four
terms of the 
strong-coupling expansion approximate the exact resonance energies
up to surprisingly small values of $g$, but both the real 
as well as the imaginary part deviate
substantially for $g \lesssim 0.025$.}
\end{center}
\end{minipage}
\end{center}
\end{figure}

{\em Method III.} We employ a strong-coupling expansion in complex
coordinates, to complement the weak-coupling 
perturbative method. Specifically, we employ the so-called Symanzik scaling 
$q \to q \, g^{-1/10}$ and rewrite the cubic Hamiltonian
into a scaled one $H_{\rm c} \to H_{\rm s}$ 
with the same eigenvalues but a fundamentally different structure, 
\begin{equation}
\label{potential_symanzik_scaling}
H_{\rm s} \, = \, 
g^{1/5} \left( H_{\ell} + \frac{q^2}{2} \, g^{-2/5} \right) \,,
\qquad
H_{\ell} =  - \frac{1}{2} \frac{\partial^2}{\partial q^2} + q^3 \,.
\end{equation}
A strong--coupling perturbation expansion can thus 
be written for each energy $E_N(g)$, which reads
\begin{equation}
\label{expansion_higher_orders}
E_N(g) \, = \, g^{1/5} \,
\sum\limits_{K = 0}^{\infty} L_{N,K} \, g^{-2K/5} \,,
\end{equation}
where $L_{N,0}$ is just equal to the $N$th 
level of $H_{\ell}$. Based on the strong-coupling
expansion, it is easy to see that a rotation angle 
$\theta = \pi/5 = 36^\circ$ is sufficient to uncover all
resonance energies of the cubic oscillator, and this angle 
is therefore chosen for all numerical calculations
reported henceforth in the current investigation.
The coefficients of the strong-coupling expansion are given in 
Table~\ref{table1}, and a comparison of the values obtained 
by method~I is made in Fig.~\ref{Fig3}.

We conclude that the numerical approach (method I) can be verified
to high accuracy
against both weak and strong-coupling expansions and is found to be 
the most convenient method to cover all ranges of the coupling
constant $g$. Furthermore, the mutual agreement of this 
numerical method with both the strong and the weak-coupling 
expansions confirms that an estimation of the numerical uncertainty 
of the results based on the apparent convergence 
of the energy levels under an increase of the size of the 
basis of states appears to be reliable. This observation means that we 
are, in principle, in a position to use the numerically determined 
resonance energies and resonance for an adiabatic time propagation
algorithm, which will be the subject of the next section of this article.

\begin{table*}[t]
\begin{center}
\begin{minipage}{16cm}
\begin{center}
\caption{\label{table1} The coefficients $L_{N,K}$ of the strong--coupling
expansion (\ref{expansion_higher_orders}) of the eigenvalues $E_N(g)$ of 
the cubic potential. For $K = 1$, the coefficients are zero. For $K = 3$,
based on numerical evidence,
we conjecture that the coefficient is independent of $N$, purely real
and equal to $1/108 = 0.009259259\dots$}
\begin{tabular}{c@{\hspace{0.6cm}}r@{\hspace{0.6cm}}r@{\hspace{0.6cm}}r}
\hline
\hline
$K$ &
\multicolumn{1}{c}{$N=0$} &
\multicolumn{1}{c}{$N=1$} &
\multicolumn{1}{c}{$N=2$} \\
\hline
0 &  $0.617~160~050 - 0.448~393~023 \, {\rm i}$ &  
     $2.193~309~731 - 1.593~532~797 \, {\rm i}$ & 
     $4.036~380~020 - 2.932~601~744 \, {\rm i}$ \\
2 & $-0.013~228~193 + 0.040~712~191 \, {\rm i}$ & 
    $-0.022~015~998 + 0.067~758~274 \, {\rm i}$ & 
    $-0.027~024~360 + 0.083~172~425 \, {\rm i}$ \\
3 & $0.009~259~259 + 0.000~000~000 \, {\rm i}$ &  
    $0.009~259~259 + 0.000~000~000 \, {\rm i}$ & 
    $0.009~259~259 + 0.000~000~000 \, {\rm i}$ \\
4 & $-0.000~294~361 - 0.000~905~951 \, {\rm i}$ & 
    $-0.000~141~177 - 0.000~434~499 \, {\rm i}$ & 
    $-0.000~118~189 - 0.000~363~747 \, {\rm i}$ \\ 
\hline
\hline
\end{tabular}
\end{center}
\end{minipage}
\end{center}
\end{table*}
%
%

%
% Figure 3 
%

%
% UNIFYING STRUCTURE AND DYNAMICS
% 
\section{UNIFYING STRUCTURE AND DYNAMICS}
\label{dynamics}

We now attempt to reconcile the structure of the 
cubic potential with its dynamics in a unified 
framework, inspired by a number of 
investigations regarding quantum dynamics formulated in complex
coordinates~\cite{Re1982arpc,MoiseyevMcCurdy,WiScWa2006,BeGaNa2007}. 
Notice that the construction of a time propagation algorithm
from complex coordinates was explicitly mentioned as a desirable
goal a rather long time ago, in Ref.~\cite{Re1982arpc}.
Note also that the time propagation of wave packets in a cubic potential 
is not completely trivial: We consider 
a particle initially at rest and located at $q(t=0) < -(3 \, \sqrt{g})^{-1}$
in the cubic potential $V(q) = \half \, q^2 + \sqrt{g}\, q^3$.
According to classical mechanics, this particle
reaches $q = -\infty$ in a finite time (as is well known), and for 
quantum mechanics, this means that 
the component of a wave packet located, loosely speaking,
to the left of the cubic well is accelerated toward $q = -\infty$
by the cubic term, consistent with the Ehrenfest theorem,
and escapes any (necessarily finite) grid in coordinate space
used for the time propagation in a finite time.
This ``escape mechanism,'' which leads to a loss of probability amplitude 
for any part of the wave packet located in a finite subinterval of
coordinate space, affords a physically intuitive explanation
for the finite decay width associated with the resonance 
eigenstates of the cubic potential.

\begin{figure}[t]
\begin{center}
\begin{minipage}{12cm}
\begin{center}
\vspace*{-0.0cm}
\includegraphics[width=1.0\linewidth,angle=0]{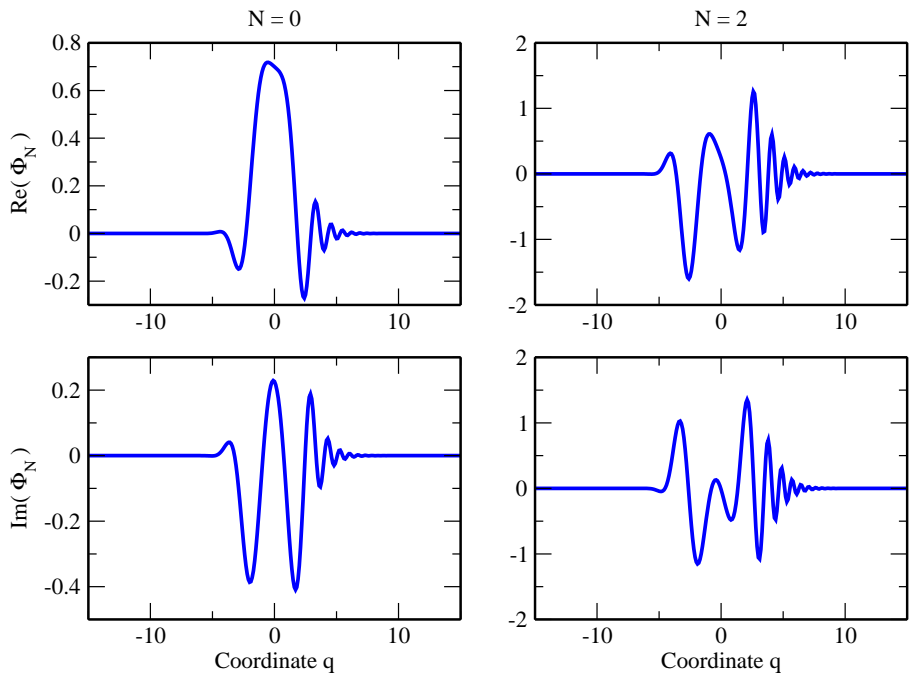}
\vspace*{-0.0cm}
\caption{\label{Fig4} (color online.) Real (top panels) and imaginary 
(bottom panels) parts of the wavefunction $\Phi_{N}$ of the cubic Hamiltonian
as a function of coordinate $q$. Results have been obtained 
for the ground state $N = 0$ and for the second excited state 
with $N = 2$. The coupling used is $\sqrt{g} = 0.1$ and the
universal complex rotation angle is $\theta = 36^\circ$.
Note that both the real and the imaginary parts of the 
ground-state wave function have nodes, but the wave function has
no zero when considered as a complex variable.}
\end{center}
\end{minipage}
\end{center}
\end{figure}

We proceed as follows. Using
a basis spanned by the standard harmonic
oscillator wavefunctions $\{ \phi_J(q) \}_{J=0}^\infty$,
we expand the cubic eigenfunctions $\Phi_N(q)$ of the
complex scaled Hamiltonian~(\ref{GENcubic_scaled}),
which fulfill $H_{\rm c}(\theta)\, \Phi_N(q) = E_N \, \Phi_N(q)$
with complex $E_N$, as follows:
\begin{equation}
\label{cubic_eigenfunctions}
\Phi_N(q) \, = \, \sum\limits_{J=0}^{\infty} c_{N,J}\, \phi_J(q) \, .
\end{equation}
Here, the complex coefficients $c_{N,J}$ are found by the diagonalization
of the complex scaled
Hamiltonian (\ref{GENcubic_scaled}). This immediately 
implies that the eigenfunctions of the (dilationally--transformed) cubic 
potential are also complex. In Fig.~\ref{Fig4}, for example, 
we display the 
real and imaginary parts of the eigenfunctions of the ground ($N$ = 0)
and excited ($N$ = 2) states of the cubic potential with
$\sqrt{g}$ = 0.1. 

After the 
evaluation of the eigenfunctions (\ref{cubic_eigenfunctions}), the 
question may arise whether these function form a complete basis set. 
In fact, such a question is not completely trivial: it is 
known~\cite{MoCeWe1978,MoFr1980,MoiseyevMcCurdy} that for very special values of 
the dilational
parameter $\theta$, an incomplete basis of the Hamiltonian can be 
obtained (see Sec.~2.5 of Ref.~\cite{MoiseyevMcCurdy}). 
In particular, it can be shown that for special, isolated
values of $\theta$, the number of linearly
independent vectors of the complex-scaled potential may be smaller
than the size of the Hamiltonian matrix. We note, however, that 
any infinitesimally small variation of $\theta$ turns the 
spectrum into a complete one~\cite{MoiseyevMcCurdy}. Below, therefore, 
we assume that the set of eigenfunctions (\ref{cubic_eigenfunctions}) form 
a complete orthonormal basis in the complex--transformed space. That is,
they fulfill the condition:
\begin{equation}
\label{orthonormalization}
\left( \Phi_N|\Phi_M \right) = \delta_{NM} \, ,
\end{equation}
where $\left( \cdot |\cdot \right)$ is the so--called $c$--inner product
introduced by N.~Moiseyev and coworkers \cite{MoCeWe1978, MoiseyevMcCurdy}:
\begin{equation}
\label{c-norm}
\left( \Phi_N|\Phi_M \right) \, = \, 
\int \Phi_N(q) \, \Phi_M(q) \, {\rm d}q \, .
\end{equation}
In order to illustrate the properties of the $c$-inner product,
we represent the Schr\"odinger equation 
as a matrix eigenvalue problem \cite{MoiseyevMcCurdy}
with a Hamiltonian matrix $\bm H$, which we denote in boldface 
notation in order to emphasize that it refers to the matrix obtained
after the complex rotation. In such a representation, 
the ``bra'' and ``ket'' eigenstates of the Hamiltonian are 
given by the ${\bm \Phi}^{\cal L}_j$ and 
${\bm \Phi}^{\cal R}_j$ row and column eigenvectors which 
satisfy the matrix equations
\begin{equation}
\label{left_vector}
   {\bm H} {\bm \Phi}^{\cal R}_j = 
   E_j {\bm \Phi}^{\cal R}_j \, ,
\end{equation}
and    
\begin{equation}
\label{right_vector}
\left( {\bm \Phi}^{\cal L}_j \right)^{\rm t} {\bm H}  =
\left( {\bm H}^{\rm t} \, {\bm \Phi}^{\cal L}_j \right)^{\rm t} =
E_j \left( {\bm \Phi}^{\cal L}_j \right)^{\rm t} \,.
\end{equation}
We recall that the eigenvalues of a matrix and its transpose are
necessarily the same (as follows from the secular equation), 
while the corresponding left and right 
eigenvectors are not necessarily the same.
Since the left eigenvectors ${\bm \Phi}^{\cal L}_j$ are
{\em eo ipso} transposed and satisfy the eigenvalue equation
of the transposed Hamiltonian, it is not necessary
to invoke complex conjugation in the definition of the inner product
(\ref{c-norm}).
When the Hamiltonian 
matrix is derived from an originally ``purely real'' Hamiltonian 
such as (\ref{GENcubic})
by complex scaling and thus symmetric (${\bm H} = {\bm H}^{\rm t}$), 
then the right and left vectors are identical,
${\bm \Phi}^{\cal L}_N = {\bm \Phi}^{\cal R}_N \equiv {\bm \Phi}_N$. 
This is the case for the cubic Hamiltonian.

We now return to the eigenfunctions (\ref{cubic_eigenfunctions}) 
of the cubic Hamiltonian. These wavefunctions, which
form a complete basis, may be utilized
for the numerical integration of the (single--particle) 
Schr\"odinger equation and, hence, for studying the dynamical 
evolution of the wave packet in the cubic potential. 
Note, however, that this packet is defined initially in the 
normal (i.e., not in the scaled) space. The time propagation, therefore, 
requires first a complex 
scaling of the initial wave packet:
\begin{equation}
\label{initial_packet_transformation}
\Psi(q, t=0) \to  \Psi(q \, {\rm e}^{{\rm i}\,\theta}, t=0) \equiv 
\Psi_{\rm c}(q, t=0)\, .
\end{equation}
After the complex scaling, the initial wave packet can be 
expanded in the basis of the functions (\ref{cubic_eigenfunctions})
\begin{equation}
\label{initial_packet_decomposition}
\Psi_{\rm c}(q, t=0) = \sum\limits_{N=0}^\infty b_N \, \Phi_N(q) \, ,
\end{equation}
where $b_N = \left( \Phi_N|\Psi_{\rm c}(t = 0 ) \right)$. 
The propagation of the wave packet $\Psi_{\rm c}(q, t)$ in the 
complex scaled coordinates is finally given by:
\begin{equation}
\label{packet_propagation}
\Psi_{\rm c}(q, t) =
\sum\limits_{N=0}^\infty \, b_N \, 
{\rm e}^{-{\rm i} \, E_{N} \, t} \,
\Phi_N(q) =
\sum\limits_{N=0}^\infty b_N \,
{\rm exp}\left(-\frac{\Gamma_N}{2} \, t\right) \,
{\rm e}^{-{\rm i} \, {\rm Re}(E_{N}) \, t} \,
\Phi_N(q) \,. 
\end{equation}
After performing the time propagation, the final function
$\Psi_{\rm c}(q, t)$ should be transformed back into the ``normal''
space, to obtain the wave function $\Psi(q, t)$.
The back-transformation proceeds by a simple inversion
of the replacement operation~(\ref{initial_packet_transformation}),
\begin{equation}
\label{backtrafo}
\Psi_{\rm c}(q, t)\to 
\Psi(q, t) \equiv  \Psi_{\rm c}(q \, {\rm e}^{-{\rm i}\,\theta}, t) \, .
\end{equation}
%

%
% Figure 4 
%
%
%

%
% Figure 5 
%
\begin{figure}[t]
\begin{center}
\begin{minipage}{12cm}
\begin{center}
\vspace*{-0.0cm}
\includegraphics[width=0.8\linewidth,angle=0, clip=]{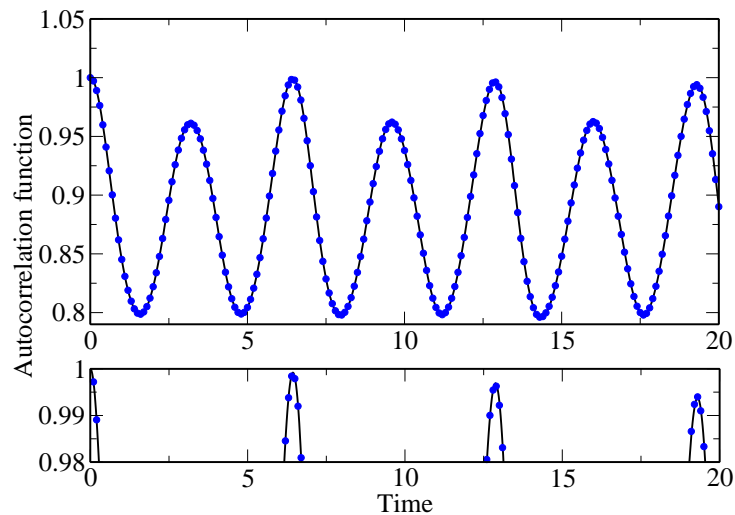}
\caption{\label{Fig5} (color online.) Time evolution of the autocorrelation 
function (\ref{ac-function}). Calculations are performed for an 
(initially) Gaussian wave packet of the form 
$\Psi(q, t = 0) = {\rm e}^{-q^2} / (\pi/2)^{1/4}$
in a cubic potential with the coupling 
constant $\sqrt{g} = 0.04$. Results obtained from 
Eq.~(\ref{packet_propagation}) are compared with those of the
well-known Crank-Nicolson approach (solid line).
The lower panel is a close-up of the upper one, where the 
range of the ordinate axis in restricted to the interval
$P(t) \in (0.98,1.0)$. The maximum of $P(t) = 0.9939$ 
near the sixth oscillation, at time $t = 19.29$, is faithfully
reproduced by the complex scaled propagation.}
\end{center}
\end{minipage}
\end{center}
\end{figure}

Using Eq.~(\ref{packet_propagation}),
we perform calculations for the time propagation of an (initially) 
Gaussian wave packet within the cubic potential with the coupling
parameter $\sqrt{g}$ = 0.04. In order to visualize the results of such 
a time propagation, we display in Fig.~\ref{Fig5}
the so--called autocorrelation function 
\begin{equation}
   \label{ac-function}
   P(t) = \left| \langle \Psi(t)|\Psi(0) \rangle \right|^2 \, ,
\end{equation}
which is commonplace for studying tunneling phenomena in 
multi--well quantum dot potentials \cite{GrDiJuHa1991}. 
As seen from Fig.~\ref{Fig5}, the wave packet, initially 
located inside the well (at $q$ = 0), performs oscillations
confined within the well. With every cycle of the oscillation, however, the 
autocorrelation function slightly decreases due to 
tunneling. We have also verified numerically that the 
decay of a wave packet corresponding to a single 
resonance eigenstate is described correctly by the back-transformation
(\ref{backtrafo}), via a comparison to a Crank-Nicolson method,
where the spatial box had to be chosen sufficiently large in the 
latter case to accommodate the spreading of the wave packet for 
larger propagation times.

We note that suitable generalizations of the method 
(\ref{packet_propagation}) can also be applied to time--dependent 
(driven) potentials with resonances, e.g.,~within the adiabatic approximation,
which is valid provided one uses sufficiently small time steps
so that the potential can be regarded as constant in time
within each time step.
Recently, the adiabatic approach has been successfully applied 
to study the propagation of the wave packet in the driven 
double--well--like potential \cite{SuLuZJJe2006}.

Up to now, we have considered the idealized case
of a decoherence-free, dissipation-free time evolution of the
metastable states in potentials which can be
approximated by a cubic Hamiltonian.
The coupling of the system to an ensemble
of harmonic oscillator modes, whose levels are
occupied according to thermal distributions, changes
the dynamics. The latter model has been studied
extensively in the literature where it is commonly referred to 
as the Caldeira--Leggett model~\cite{CaLe1981},
and it has been discussed at length in a number of research
articles and books~\cite{CaLe1983,GrWeHa1984,%
GrWe1984,LaOv1984,GrOlWe1987,CaIm1998,ClMaCl1988,In1997,%
We1999}; therefore its details are relegated to the 
Appendix~\ref{dissipation}. 
The model describes dissipation and the corresponding
quantum decoherence, and, in general, leads to
multiplicative corrections to the
decay widths of the resonance eigenstates, of the form
\begin{equation}
\label{incorp}
\Gamma_N \to \Gamma_N \, (1 + \delta_N)
\end{equation}
for the decay widths. The corrections $\delta_N$, initially,
could be assumed as affecting only the 
structure of the resonance energies, but
using our time propagation, it is easy to include them into
the dynamics as well, by simply applying the 
replacement (\ref{incorp}) to the $\Gamma_N$
which are present in Eq.~(\ref{packet_propagation}). 

%
% Summary and Outlook
%
\section{SUMMARY AND OUTLOOK}
\label{summary}

In this paper, we have illustrated a certain duality of the cubic 
anharmonic oscillator; for purely imaginary coupling, the eigenenergies are
real, whereas for real coupling, the resonance energies are complex
(see Secs.~\ref{theoryA} and~\ref{theoryB}).
Moreover, for imaginary coupling, there are no instanton
configurations to consider, and the quantization condition involves
only the ``perturbative $B$ function,'' suitably resummed, 
and is of the plain Bohr--Sommerfeld type [see Eq.~(\ref{BSQC})]. 
By contrast, for 
real coupling, there are instanton configurations present, and these
manifest themselves in nonvanishing decay widths of the state
and in a modified quantization condition with allowance for the 
instanton configurations. The modified quantization is not 
obtained here, but we lay the groundwork for its construction.
[Specifically, the leading term of the ``instanton $A$ function'' is given 
in Eq.~(\ref{instA}).]

In other words, even anharmonic oscillators (quartic, sextic, etc.), 
for real coupling, are described 
by a plain Bohr--Sommerfeld quantization condition, whereas for negative 
coupling, they give rise to instantons. The structure of an 
even oscillator for negative coupling is thus in a certain sense
analogous to an 
odd oscillator for real coupling $\sqrt{g} \in \mathbbm{R}$, and 
the presence of instantons and the corresponding 
nontrivial saddle points of the Euclidean action demand a modification
of the Bohr--Sommerfeld quantization condition.
By contrast, an even oscillator for positive coupling parameter is analogous 
to an odd oscillator for purely imaginary coupling $\sqrt{g} = \ii \,
\beta$ and is described by a plain Bohr--Sommerfeld quantization condition
and completely characterized by Rayleigh--Schr\"{o}dinger 
perturbation theory.

For the determination of the resonance energies of the 
cubic potential (positive coupling, see Sec.~\ref{theoryB}), 
three different methods have been compared.
Method I (numerical approach in complex coordinates) is found 
to be more universal in applicability than method II (resummed
weak-coupling expansion) and method III (strong-coupling expansion),
although good agreement is observed in the specified domains 
of applicability for each of the latter two methods. 
In any case, the numerical approach 
provides us with resonance energies and corresponding 
wave functions in complex space which can be used in order
to unify the analysis of the structure and of the dynamics 
of quantal particles. Hereby, the immediate physical interpretation
of the decay widths can be illustrated in a particularly 
clear way by investigating the time propagation of wave packets.

As shown in Sec.~\ref{dynamics},
the complex rotated eigenfunctions evolve in time according
to complex resonance energies which describe the quantum
tunneling and the decay of the wave packets. When transformed
back into real space, the algorithm leads to results which 
are in agreement with traditional quantum dynamics solvers
like the Crank-Nicolson method. The structure and the 
dynamics of the potentials are thus obtained in a single,
unified method which allows for an inclusion of 
modifications to the widths of the resonances due to dissipation
(coupling to the environment).
For large propagation times, our algorithm is numerically stable, because 
the time evolution of the resonance eigenstates given by 
Eq.~(\ref{packet_propagation}) remains valid for the entire 
coordinate space. The ``escape'' of the wave packet
toward $q \to -\infty$, which is more pronounced for the cubic 
potential than, e.g., for the linear ponderomotive potential known from laser
physics, is thus automatically incorporated into our algorithm.
Moreover, slight modifications of the structural properties 
of the resonances due to additional perturbations (e.g.,
couplings to an environment) can easily be incorporated into the 
dynamics, as exemplified by Eq.~(\ref{incorp}) and discussed in more
detail in Appendix~\ref{dissipation} below.

%
% Acknowledgments
%
\section{ACKNOWLEDGMENTS}

U.D.J. acknowledges support from the Deutsche
Forschungsgemeinschaft (Heisenberg program) and helpful
discussions with Professor O.~Nachtmann.

\appendix
\section{DISSIPATION IN QUANTUM TUNNELING}
\label{dissipation}

Recently, a rather large number of theoretical 
\cite{StEtAl2003,MaEtAl2003prb,BuEtAl2003prl,GeCl2005} 
as well as experimental 
\cite{MaEtAl2002,BeEtAl2003sc,BeEtAl2003prb,XuEtAl2005,KaEtAl2006} 
studies have been devoted to the 
current-biased Josephson junction since its low-lying energy levels 
can be used to implement a solid-state quantum bit (qubit). This qubit 
circuit might be considered as a conceivable
basis of a scalable quantum computer. 
The theoretical analysis of the structure and dynamics of Josephson
junction qubits can be traced back to the cubic potential,
at least in the limit of a dissipation-free, or decoherence-free
environment, which is an idealized scenario that 
is nevertheless pursued in the construction 
of quantum computing devices.  
The measurement of the qubit state utilizes the escape
from the cubic potential via tunneling and thus it is 
important to have the dynamical aspects of related
potential under control.

When we work with a plain cubic Hamiltonian, it is clear
that the analysis applies only to the idealized case
of a decoherence-free, dissipation-free time evolution of 
metastable states in potentials which can be 
approximated by a cubic Hamiltonian.
The coupling of the system to an ensemble 
of harmonic oscillator modes, whose levels are
occupied according to thermal distributions, changes
the dynamics and can be incorporated 
approximately into our algorithm by the 
replacement (\ref{incorp}). The so-called 
Caldeira--Leggett model has been studied
extensively in the literature~\cite{CaLe1981,CaLe1983,GrWeHa1984,%
GrWe1984,LaOv1984,GrOlWe1987,CaIm1998,ClMaCl1988,In1997,%
We1999} and describes dissipation and the corresponding 
quantum decoherence.
Ideally, of course, one has no dissipation in a device
used for quantum computing, and this 
scenario has been the focus of the current investigation.
It might be worth emphasizing that 
our approach allows for a full access to the quantum dynamics
in the limit of vanishing dissipative terms, whereas the approach
originally outlined in Refs.~\cite{CaLe1981,CaLe1983,GrWeHa1984,%
GrWe1984,LaOv1984} can be used to investigate the rate at which the
system loses information about coherent quantum states
in the presence of dissipation, but does not lead to a general description
of the full quantum dynamics in time-dependent potentials. 

In general, the coupling to other degrees of freedom leads to 
the phenomenon of dissipation where the energy is transferred irreversibly 
from the system under consideration to the environment. 
In the work by A.~O.~Caldeira and A.~J.~Leggett \cite{CaLe1981}, 
the so--called ``system--plus--bath'' model has been introduced
to describe dissipation. Within this model, the bath 
(environment) is considered to be 
representable as a set of harmonic oscillators interacting linearly with 
the system under consideration (i.e., with our
cubic potential). The total Hamiltonian for such a system can be written as
\begin{eqnarray}
\label{cubic_in_bath}
H^{\rm diss}_{\rm c} & = & 
\frac{p^2}{2}  + \frac{1}{2} q^2 + 
\sqrt{g} \, q^3 \nonumber \\
& & \hspace*{-1.5cm} + \sum\limits_{i = 1}^N \left( \frac{p_i^2}{2m_i}
+ \frac{m_i \omega_i^2}{2} x_i^2 - q c_i x_i + q^2 
\frac{c_i^2}{2 m_i \omega^2_i }\right) \, ,
\end{eqnarray}
where the first line displays ``environmentally decoupled'' system, 
i.e.~the cubic anharmonic
oscillator in our case. The sum over $i$ in Eq.~(\ref{cubic_in_bath})
contains the Hamiltonians for a set of $N$ harmonic oscillators
which are bilinearly coupled with 
strength $c_i$ to the system. Finally, the last 
term represents a potential renormalization term \cite{In1997,We1999}.

For the practical application of Eq.~(\ref{cubic_in_bath}),
it is necessary to 
eliminate the external degrees of freedom, i.e.~the bath modes
labeled $1,\dots,N$. Since these are just harmonic
oscillator modes, we can easily solve the 
equations of motion for the external degrees of freedom.
In a classical framework, we can derive the
equation of motion for the cubic anharmonic oscillator in the 
bath as\cite{In1997,We1999}
\begin{equation}
   \label{equation_motion}
   \ddot{q} + \int\limits_{0}^t {\rm d}t' \gamma(t-t') \dot{q}(t') +
   q + 3 \sqrt{g} q^2 = \xi(t) \, .
\end{equation}
Here, the damping kernel $\gamma$ is given by
\begin{equation}
   \label{damping}
   \gamma(t) = \sum\limits_{i = 1}^N \frac{c_i^2}{m_i \omega^2_i}
   \cos(\omega_i t) \, ,
\end{equation}
and the fluctuating acceleration
\begin{eqnarray}
   \label{fluctuation}
   \xi(t) &=& -\gamma(t) \, q(0) \nonumber \\
   &+& \sum\limits_{i = 1}^N c_i 
   \left( x_i(0) \cos(\omega_i t) + \frac{p_i(0)}{m_i \omega_i}
   \sin(\omega_i t) \right) \, 
\end{eqnarray}
contains the initial conditions for the bath modes.

On a fully quantum level,
it is convenient to describe the decay of the metastable 
state coupled to environment within the path integral formalism,
because harmonic oscillator modes can easily be integrated out. 
The analysis of 
dissipative quantum tunneling is 
naturally performed in the Euclidean 
space (see Refs.~\cite{In1997, We1999} for more details). 
That is, after a suitable Wick rotation, the 
imaginary--time path integral provides a representation 
of the equilibrium density matrix which can be represented as
\begin{equation}
   \label{equilibrium_dm}
   \rho_\beta(q_i, q_f) = \frac{1}{Z_{\beta}} 
   \int\limits_{q(0) = q_i}^{q(\beta) = q_f} 
   \mathcal{D}q \, \exp \left( - S^{E}[q] \right) \,,
\end{equation}
where $Z_{\beta}$ is the partition function which here
acts as a normalization prefactor, and 
$\beta = 1/(k_B T)$ is the
imaginary or thermal time (we set $\hbar = 1$). The paths in the integral 
(\ref{equilibrium_dm}) are weighted with a phase factor that contains 
the effective Euclidean action for the anharmonic oscillator
obtained after integrating out the modes of thermal
bath~\cite{CaLe1981,CaLe1983,In1997,We1999} and reads
\begin{eqnarray}
   \label{action}
   S^{E}[q] &=& \int\limits_{0}^{\beta} d\tau 
   \left( \frac{p^2}{2} + \frac{1}{2} q^2 + 
   \sqrt{g} \, q^3 \right) \nonumber \\ 
   &+& \frac{1}{2}  \int\limits_{0}^{\beta} d\tau 
   \int\limits_{0}^{\beta} d\tau' k (\tau - \tau') \,
   q(\tau) \, q(\tau') \, .
\end{eqnarray}
The first term of this action is just a standard 
Euclidean action of the cubic Hamiltonian (\ref{GENcubic}),
while the second term describes the frictional influence of the environment.
The temperature-dependent, nonlocal (in time) 
damping kernel $k(\tau - \tau')$ may be expressed as a Fourier series
(see, e.g., Refs.~\cite{GrOlWe1987,We1999})
\begin{equation}
\label{kernel}
k(t) = \frac{1}{\beta} 
\sum\limits_{n = -\infty}^{+\infty} \xi_n \exp({\rm i}\, \nu_n \, t) \, ,
\end{equation}
where the $\nu_n = 2 \pi n/ \beta$ are bosonic Matsubara 
frequencies. Finally, we have $\xi_n =  |\nu_n| \,
\hat{\gamma}(|\nu_n|)$, where $\hat{\gamma}$ denotes the Laplace 
transform of the damping kernel (\ref{damping}).

Based on the path integral approach (\ref{equilibrium_dm})--(\ref{kernel}),
the quantum--mechanical decay of the cubic oscillator 
under the influence of a heat bath environment has been studied. 
In particular, explicit results have been obtained which helped to understand
how the tunneling rate depends on the temperature of the environment,
in the presence of medium and strong damping 
terms~\cite{GrWeHa1984,GrOlWe1987,We1999}.

\end{document}